\documentclass[preprintnumbers,amsmath,amssymb,floatfix,12pt,prd,superscriptaddress,nofootinbib]{revtex4}
\usepackage{graphicx}
\usepackage{epsfig}
\usepackage{bm}
\usepackage{amsfonts}
\usepackage{subfigure}

\def\bea{\begin{eqnarray}}
\def\eea{\end{eqnarray}}

\begin{document}
\begin{center}
\LARGE {\bf Warm-viscous inflation  model  on the brane in the light of BICEP2}
\end{center}
\begin{center}
{M. R. Setare $^{a}$\footnote{E-mail: rezakord@ipm.ir
}\hspace{1mm} ,
V. Kamali $^{b}$\footnote{E-mail: vkamali1362@gmail.com, Vkamali@basu.ac.ir}\hspace{1.5mm} \\
 $^a$ {\small {\em Department of Science, Campus of Bijar, University of Kurdistan.\\
  Bijar , Iran.}}\hspace{1.5mm}\\
$^b$ {\small {\em  Department of Physics, Faculty of Science,\\
Bu-Ali Sina University, Hamedan, 65178, Iran}}}\\

\end{center}


\begin{center}
{\bf{Abstract}}\\
In the present work warm inflationary universe model with viscous pressure  on the brane in high-dissipation  regime is studied. We derive
 a condition which is required for this model to be realizable in slow-roll approximation. We also present analytic expressions for density perturbation and amplitude of tensor perturbation in longitudinal gauge. General expressions of tensor-to-scalar ratio, scalar spectral index and its running are obtained.
We develop our model by using chaotic
potential, the characteristics of this model are calculated for two specific cases: 1- Dissipative parameter $\Gamma$ and bulk viscous parameter $\zeta$ are constant parameters.
2- Dissipative parameter  as a function of scalar field $\phi$ and bulk viscous parameter
as a function of radiation-matter mixture  energy density $\rho$. The parameters of the model are restricted by
 WMAP9, Planck  and BICEP2 observational data.
 \end{center}

\newpage

\section{Introduction}
 Big Bang model has many long-standing problems (horizon,
flatness,...). These problems will find explanations in a framework of
inflationary universe models \cite{1-i}. Scalar field as a source
of inflation provides a causal interpretation of the origin of
the distribution of large scale structure (LSS), and also observed anisotropy
of cosmological microwave background (CMB) \cite{6,planck,BICEP2}. Standard
models for inflationary universe are divided
into two regimes, slow-roll and reheating epochs. In the slow-roll
period, when the kinetic energy is small compared to the potential
energy terms, the inflation period appears. In this period, all interactions between the scalar fields
(inflatons) and  other fields (radiation, matter...) are neglected.  Subsequently, in the reheating epoch, the kinetic energy
is comparable to the potential energy that causes the inflaton
oscillate  around  the minimum of the potential while losing their
energy to other fields present in the theory. After this period, the universe is filled with radiation. \\ In warm inflation scenario, the radiation production
occurs during the slow-roll inflation period and the
reheating is avoided \cite{3}. Thermal fluctuations may be
produced during warm inflationary epoch. These fluctuations could play a
dominant role to produce initial fluctuations which are necessary
for the Large-Scale Structure (LSS) formation. In this model, density
fluctuation arises from thermal rather than quantum fluctuation
\cite{3-i}. Warm inflationary period ends when the universe stops
inflating. After this period, the universe enters in the radiation
phase smoothly \cite{3}. Finally, remaining inflatons or dominant
radiation fields create matter components of the universe. Some extensions of this model are found in Ref.\cite{new}.\\
In warm inflation models, for simplicity, particles which are created by the inflaton decay are considered as massless particles (or radiation). Existence of massive particles in the inflationary fluid model as a new model of inflation was considered in Ref.\cite{4-i}.
Perturbation parameters of this model were obtained in Ref.\cite{2-ne}.
In this scenario the existence of massive particles may altere the dynamic of the inflationary universe models by modification of the fluid pressure.
Decay of the massive particles within the fluid is an entropy-producing scalar phenomenon. On the other hand,
''bulk viscous pressure'' has entropy-producing property.
The decay of particles  may be considered by a bulk viscous pressure $\Pi=-3\zeta H$, where $H$ is
Hubble parameter and $\zeta$ is phenomenological coefficient of bulk viscosity \cite{3-ne}.
This coefficient is positive-definite by the second law of thermodynamics and in general depends on the energy density of the fluid.\\

We may live on a brane which is embedded in a higher
dimensional universe. This realization has significant implications to cosmology
\cite{1-f}. In this scenario, which is motivated by string theory, gravity (closed string modes)
can propagate in the bulk, while the standard model of particles (matter fields which are related to open string modes) is confined
to the lower-dimensional brane \cite{2-f}. $4d$ Einstein's equation projected onto the brane has been found in Ref.\cite{3-f}. Friedmann equation and the equations of linear perturbation theory \cite{4-f} may be modified by these projections. We would like to study the warm  inflation model with viscous pressure  in the Randall-Sundrum braneworld  model \cite{Ra}.
We will study the linear cosmological perturbations theory for warm inflation model with viscous pressure on the brane.
In the Randall-Sundrum cosmological model, the idea that the universe is in a 3d-brane within an extra-dimensional bulk spacetime, is successfully consolidated by the issue of inflation of a single brane in an AdS bulk \cite{Ra}.
Einstein's equation in the Randall-Sundrum braneworld with cosmological constant as a source and matter fields confined to 3-brane may be projected on to the brane as a following equation \cite{3-f}
\begin{equation}\label{1}
G_{\mu\nu}=-\Lambda_4 g_{\mu\nu}+(\frac{8\pi}{M_4^2})T_{\mu\nu}+(\frac{8\pi}{M_5^2})^2\pi_{\mu\nu}-E_{\mu\nu},
\end{equation}
where $M_4$ and $M_5$ are Planck scales in 4 and 5 dimensions, respectively, $E_{\mu\nu}$ is a projection of 5d weyl tensor, $T_{\mu\nu}$ is energy density tensor on the brane and $\pi_{\mu\nu}$ is a tensor  quadratic in $T_{\mu\nu}$. Effective cosmological constant $\Lambda_4$ on the brane in terms of 3-brane tension $\lambda$ and $5d$ cosmological constant $\Lambda$ is given by
\begin{equation}\label{}
\nonumber
\Lambda_4=\frac{4\pi}{M_5^3}(\Lambda+\frac{4\pi}{3M_5^3}\lambda^2),
\end{equation}
and $4d$ Planck scale is determined by $5d$ Planck scale as
\begin{equation}\label{}
\nonumber
M_4=\sqrt{\frac{3}{4\pi}}(\frac{M_5^2}{\sqrt{\lambda}})M_5.
\end{equation}
In spatially flat Friedmann-Robertson-Walker (FRW) model, Friedmann equation, from equation (\ref{1}), has  the following form \cite{3-f}
\begin{equation}\label{}
\nonumber
H^2=\frac{\Lambda_4}{3}+(\frac{8\pi}{3M_4^2})\rho_T+(\frac{4\pi}{3M_5^3})\rho_T^2+\frac{\varepsilon}{a^4},
\end{equation}
where $a$ is scale factor of the model and $\rho_T$ is  total energy density on the brane. The last term in the above equation denotes the influence of the bulk gravitons on the brane, where $\varepsilon$ is an integration constant which arises from Weyl tensor $E_{\mu\nu}$. This term may be rapidly diluted once the inflation begins. The projected Weyl tensor term in the effective Einstein equation may be neglected. Therefor, this term do not give the significant contributions to the observable perturbation parameters. It is also assumed  that  the $\Lambda_4$ is negligible in the early universe. The Friedmann equation is reduced to
\begin{equation}\label{}
\nonumber
H^2=\frac{8\pi}{3M_4^2}\rho_T(1+\frac{\rho_T}{2\lambda}).
\end{equation}
where we have used natural units ($c=\hbar=1$).

Here, our goal is to investigate the warm inflation model with viscous pressure in the brane scenario, where  the total energy density $\rho_{T}=\rho_{\phi}+\rho_{\gamma}$ is found on the brane \cite{5-f,6-f} ($\rho_{\phi}$ is energy density of the scalar field and $\rho_{\gamma}$ is energy density of the matter-radiation fluid.). The  Friedmann equation for our model has the form
\begin{equation}\label{2}
H^2=\frac{8\pi}{3M_4^2}(\rho_{\phi}+\rho_{\gamma})(1+\frac{\rho_{\phi}+\rho_{\gamma}}{2\lambda}).
\end{equation}
Cosmological perturbations of warm inflation model (on the brane) have been studied in Ref.\cite{9-f} (Ref.\cite{6-f}).
Warm tachyon inflationary universe model  has been studied in
Ref.\cite{1-m}, also warm inflation with viscous pressure have been studied in Ref.\cite{2-ne} . To the best of our knowledge, a model for  warm inflation with viscous pressure on the brane
has not been yet considered.
In the present work, warm
inspired inflation with viscous pressure on the brane will be studied  by using
the above modified Friedmann
equation. The paper is organized as: In next section, we will
describe warm inflationary universe model with viscous pressure in the
brane scenario. In section (3), we obtain  perturbation parameters
for our model. In section (4), we study our model  using the
chaotic potential in high dissipative regime and high energy limit. Finally in
section (5), we close by some concluding  remarks.

\section{The model}
We consider warm-inflationary model on the brane in the spatially flat FRW universe scenario which is filled with self-interacting inflaton field $\phi$ and an imperfect fluid. Scalar field $\phi$ has energy density and pressure with the forms, $\rho_{\phi}=\frac{1}{2}\dot{\phi}^2+V(\phi),$ and $P_{\phi}=\frac{1}{2}\dot{\phi}^2-V(\phi),$ respectively. The imperfect fluid is a mixture of matter and radiation of adiabatic index $\gamma$ which has an energy density $\rho_{\gamma}=Ts(\phi,T)$ ($T$ is temperature  and $s$ is entropy density of the imperfect fluid.) and pressure $P+\Pi$ \cite{Villanueva}. The dynamic of this model is given by modified Friedmann equation
\begin{eqnarray}\label{3}
3H^2=(\rho_{\phi}+\rho_{\gamma})(1+\frac{\rho_{\phi}+\rho_{\gamma}}{2\lambda})~~~~~~~~~~~~~~~~~~\\
\nonumber
=(\frac{1}{2}\dot{\phi}^2+V(\phi)+\rho_{\gamma})
(1+\frac{\frac{1}{2}\dot{\phi}^2+V(\phi)+\rho_{\gamma}}{2\lambda}),
\end{eqnarray}
and conservation equations for  inflaton field and imperfect fluid, which are connected by the dissipation term $\Gamma\dot{\phi}^2$ ($\Gamma$ is dissipation coefficient)
\begin{eqnarray}\label{4}
\dot{\rho}_{\phi}+3H(\rho_{\phi}+P_{\phi})=-\Gamma\dot{\phi}^2\Rightarrow \ddot{\phi}+3H\dot{\phi}+V'=-\Gamma\dot{\phi},
\end{eqnarray}
and
\begin{eqnarray}\label{5}
\dot{\rho}_{\gamma}+3H(\rho_{\gamma}+P+\Pi)=\dot{\rho}_{\gamma}+3H(\gamma\rho_{\gamma}+\Pi)=\Gamma\dot{\phi}^2,
\end{eqnarray}
where $V'=\frac{dV}{d\phi},$ ~$P=(\gamma-1)\rho_{\gamma},$ and $\frac{8\pi}{M_4^2}=1$.
Dissipation term denotes the inflaton decay into the imperfect fluid in the inflationary epoch.

We would like to express the evolution equation (\ref{5})  in terms of entropy
density $s(\phi,T)$. This parameter is defined by a thermodynamical relation \cite{mm-1}
\begin{equation}\label{}
s(\phi,T)=-\frac{\partial f}{\partial T}=-\frac{\partial V}{\partial T}.
\end{equation}
$f$ is Helmholtz free energy which is defined by
\begin{equation}\label{}
f=\rho_{T}-Ts=\frac{1}{2}\dot{\phi}^2+V(\phi)+\rho_{\gamma}-Ts.
\end{equation}
Free energy $f$  is dominated by the thermodynamical potential $V(\phi, T)$ in slow-roll limit \cite{Villanueva}. The total energy density and total pressure are given by
\begin{eqnarray}\label{}
\rho_T=\frac{1}{2}\dot{\phi}^2+V(\phi)+Ts,~~~~~~~~~~~~~~~~\\
\nonumber
P_T=\frac{1}{2}\dot{\phi}^2-V(\phi)+(\gamma-1)Ts+\Pi.
\end{eqnarray}
The viscous pressure for an expanding universe is negative ($\Pi=-3\zeta H$), therefore this term acts to decrease the total pressure. Using Eq.(\ref{5}), we can find the entropy density evolution for our model as
\begin{eqnarray}\label{nn-1}
T\dot{s}+3H(\gamma Ts+\Pi)=\Gamma\dot{\phi}^2.
\end{eqnarray}
In the above equation, it is assumed that $\dot{T}$ is negligible.
For a quasi-equilibrium high temperature thermal bath as an inflationary fluid, we have $\gamma=\frac{4}{3}$. The bulk viscosity effects may be read from above equation. Thus bulk viscous pressure $\Pi$ as a negative quantity, enhances the source of entropy density on the RHS of the  evolution equation (\ref{nn-1}). Therefore,  energy density of radiation  and entropy density  increase by the bulk viscosity pressure $\Pi$ (see FIG.1 and FIG.2).

During the inflationary epoch, the energy density of inflaton field $\phi$ is the order of potential, i.e. $\rho_{\phi}\sim V(\phi),$ and dominates over the energy density of imperfect fluid, i.e. $\rho_{\phi}>\rho_{\gamma}$, this limit is called stable regime \cite{mm-1}. In slow-roll limit, we have $\dot{\phi}^2\ll V(\phi),$ and $\ddot{\phi}\ll(3H+\Gamma)\dot{\phi}$ \cite{3}. When the decay of the inflaton to imperfect fluid is quasi-stable, we have $\dot{\rho}\ll 3H(\gamma\rho_{\gamma}+\Pi),$ and $\dot{\rho}_{\gamma}\ll\Gamma\dot{\phi}^2$. Therefore the dynamic equations (\ref{3}), (\ref{4}) and (\ref{5}), in slow-roll and quasi-stable limits, are reduced to
\begin{equation}\label{6}
H^2=\frac{1}{3}V(1+\frac{V}{2\lambda}),
\end{equation}

\begin{equation}\label{7}
\dot{\phi}=-\frac{V'}{\Gamma+3H},
\end{equation}
and
\begin{equation}\label{8}
\rho_{\gamma}=\frac{r\dot{\phi}^2-\Pi}{\gamma}=\frac{1}{\gamma}(\frac{r}{3(1+r)^2}(\frac{V'}{V})^2\frac{V}{1+\frac{V}{2\lambda}}-\Pi),
\end{equation}
where $r=\frac{\Gamma}{3H}$. In the present work, the analysis is restricted in high dissipative regime, i.e. $r\gg 1,$ where the dissipation coefficient $\Gamma$ is much greater than $3H$. The reason of this choice is as following. In weak dissipative regime, i.e. $r\ll 1$, the expansion of the universe in the inflationary era disperses the decay of the inflaton. There is a little  chance for interaction between the sectors of the inflationary fluid, therefore  we do not have non-negligible bulk viscosity. Warm inflation in high and weak dissipative regimes for a non-viscous warm inflation model have been studied in Refs. \cite{3} and \cite{1-ne}, respectively. Dissipation parameter $\Gamma$ may be a constant parameter or a positive function of inflaton $\phi,$ by the second law of thermodynamics. There are some specific forms for the dissipative coefficient, with the most common which are found in the literatures being the $\Gamma\sim T^3$ form \cite{mm-1},\cite{2nn},\cite{3nn},\cite{4nn}. In some works $\Gamma$ and potential of the inflaton have  same forms \cite{1-m}. In Ref.\cite{2-ne}, perturbation parameters for warm inflationary model with viscous pressure have been obtained where $\Gamma=\Gamma(\phi)=V(\phi),$ and $\Gamma=\Gamma_0=const$. In this work we will study the warm-inflation model with viscous on the brane in high dissipative regime. \\
We introduce the slow-roll parameters $\epsilon$ and $\eta$ for our model as
\begin{equation}\label{9}
\epsilon=-\frac{\dot{H}}{H^2}=\frac{1}{2(1+r)}(\frac{V'}{V})^2\frac{1+\frac{V}{\lambda}}{(1+\frac{V}{2\lambda})^2},
\end{equation}
and
\begin{equation}\label{10}
\eta=-\frac{\ddot{H}}{H^2}=-\epsilon(1+\frac{V}{\lambda})^{-2}+\frac{1}{1+r}(\frac{V''}{V})(1+\frac{V}{2\lambda})^{-1},
\end{equation}
respectively. Using Eqs.(\ref{7}), (\ref{8}) and (\ref{9}) in slow-roll limit, we find a relation between the energy density of inflaton field and the energy density of imperfect fluid
\begin{eqnarray}\label{11}
\gamma\rho_{\gamma}+\Pi=\frac{2r\epsilon}{3(1+r)}\frac{1+\frac{V}{2\lambda}}{1+\frac{V}{\lambda}}\rho_{\phi}.
\end{eqnarray}
By using the condition of inflationary epoch, i.e. $\ddot{a}>1$ or equivalently $\epsilon<1$, and the above relation, we denote that warm inflation epoch with viscus pressure on the brane could takes place when
\begin{eqnarray}\label{12}
\rho_{\phi}>\frac{3(1+r)}{2r}\frac{1+\frac{V}{\lambda}}{ 1+\frac{V}{2\lambda}}(\gamma\rho_{\gamma}+\Pi).
\end{eqnarray}
Warm inflation epoch comes to close when, $\epsilon\simeq 1$
\begin{equation}\label{13}
\rho_{\phi}=\frac{3(1+r)}{2r}\frac{1+\frac{V}{\lambda}}{ 1+\frac{V}{2\lambda}}(\gamma\rho_{\gamma}+\Pi).
\end{equation}
The number of e-folds at the end of inflation is given by
\begin{equation}\label{14}
N(\phi)=-\int_{\phi_i}^{\phi_f}\frac{V}{V'}(1+\frac{V}{2\lambda})(1+r)d\phi,
\end{equation}
or equivalently
\begin{eqnarray}\label{15}
N(\phi)=-\int_{V_i}^{V_f}\frac{V}{V'^2}(1+\frac{V}{2\lambda})(1+r)dV.
\end{eqnarray}
where the subscripts $i$ and $f$ denote begining and end of the inflation period, respectively.
\section{Perturbation}
In this section we will study inhomogeneous perturbations of the FRW background. These perturbations in the longitudinal gauge, may be described by the perturbed FRW metric
\begin{equation}\label{16}
ds^2=(1+2\Phi)dt^2-a^2(t)(1-2\Psi)\delta_{ij}dx^idx^j,
\end{equation}
where $\Phi$ and $\Psi$ are gauge-invariant metric perturbation variables \cite{7-f}. All perturbed quantities have a spatial sector of the form $e^{i\mathbf{kx}},$ where $k$ is the wave number. Perturbed Einstein field equations in momentum space have only temporal parts
\begin{equation}\label{}
\nonumber
\Phi=\Psi,
\end{equation}

\begin{equation}\label{17}
\dot{\Phi}+H\Phi=\frac{1}{2}[-\frac{(\gamma\rho+\Pi)av}{k}+\dot{\phi}\delta\phi][1+\frac{1}{\lambda}(\frac{\dot{\phi}^2}{2}+V(\phi)+\rho_{\gamma})],
\end{equation}

\begin{eqnarray}\label{18}
(\ddot{\delta\phi})+[3H+\Gamma](\dot{\delta\phi})+[\frac{k^2}{a^2}+V''+\dot{\phi}\Gamma']\delta\phi=4\dot{\phi}\dot{\Phi}-[\dot{\Phi}\Gamma+2V']\Phi,
\end{eqnarray}

\begin{eqnarray}\label{19}
(\dot{\delta\rho}_{\gamma})+3\gamma H\delta\rho_{\gamma}+ka(\gamma\rho_{\gamma}+\Pi)v+3(\gamma\rho_{\gamma}+\Pi)\dot{\Phi}\\
\nonumber
-\dot{\phi}^2\Gamma'\delta\phi-\Gamma\dot{\phi}[2(\dot{\delta\phi})+\dot{\phi}\Phi]=0,~~~~~~~~~~~~~~
\end{eqnarray}
and
\begin{equation}\label{20}
\dot{v}+4Hv+\frac{k}{a}[\Phi+\frac{\delta P}{\rho_{\gamma}+P}+\frac{\Gamma\dot{\phi}}{\rho_{\gamma}+P}\delta\phi]=0,
\end{equation}
where
\begin{equation}\label{}
\nonumber
\delta P=(\gamma-1)\delta\rho_{\gamma}+\delta\Pi~~~~~~~\delta\Pi=\Pi[\frac{\zeta_{,\rho_{\gamma}}}{\zeta}\delta\rho_{\gamma}+\Phi+\frac{\dot{\Phi}}{H}].
\end{equation}
The above equations are obtained for Fourier components $e^{i\mathbf{kx}},$ where the subscript $k$ is omitted. $v$ in the above set of equations is given by the decomposition of the velocity field ($\delta u_j=-\frac{iak_J}{k}ve^{i\mathbf{kx}}, j=1,2,3$) \cite{7-f}.

Note that the effect of the bulk (extra-dimension) to the perturbed projected Einstein field equation on the brane may be found in Eq.(\ref{17}).
We will describe the non-decreasing adiabatic and isocurvature modes of our model on super-horizon scales, i.e. $k\ll aH$. In this limit, a complete set of perturbation equations on the brane have been obtained. Therefore, the perturbation variables along the extra-dimensions in the bulk could not have any contribution to the perturbation equations on the super-horizon scales (see for example \cite{6-f}, \cite{nn-1}.). We could see the same approach for non-viscous warm inflation model on the brane in Ref.\cite{6-f}.

Warm inflation model with viscous pressure may be considered as a hybrid-like inflationary model where the inflaton field interacts with the imperfect fluid \cite{9-f}, \cite{8-f}. Entropy perturbation may be related to dissipation term \cite{10-f}. During inflationary era with slow-roll approximation for non-decreasing adiabatic modes on large scale limit, i.e. $k\ll aH,$ it is  assumed that the perturbed quantities could not vary strongly. So, we have $\dot{\Phi}\ll H\Phi$, $(\ddot{\delta\phi})\ll(\Gamma+3H)(\dot{\delta\phi})$, $(\dot{\delta\rho})\ll\delta\rho,$ and $\dot{v}\ll 4Hv$. In the slow-roll limit and by using  the above limitations, the set of perturbed equations are reduced to
\begin{equation}\label{21}
\Phi=\frac{1}{2H}[-\frac{\gamma\rho_{\gamma}+\Pi}{k}av+\dot{\phi}\delta\phi](1+\frac{V}{\lambda}),
\end{equation}

\begin{equation}\label{22}
(3H+\Gamma)(\dot{\delta\phi})+[V''+\dot{\phi}\Gamma']\delta\phi\simeq-[\dot{\phi}+2V']\Phi,
\end{equation}

\begin{equation}\label{23}
\delta\rho_{\gamma}\simeq\frac{\dot{\phi}^2}{3\gamma H}[\Gamma'\delta\phi+\Gamma\Phi],
\end{equation}
and
\begin{eqnarray}\label{24}
v\simeq-\frac{k}{4aH}(\Phi+\frac{(\gamma-1)\delta\rho_{\gamma}+\delta\Pi}{\gamma\rho_{\gamma}+\Pi}+\frac{\Gamma\dot{\phi}}{\gamma\rho_{\gamma}+\Pi}\delta\phi).
\end{eqnarray}
Using Eqs.(\ref{21}), (\ref{23}) and (\ref{24}), the perturbation variable $\Phi$ is determined
\begin{eqnarray}\label{25}
\Phi\simeq\frac{\dot{\phi}}{2H}\frac{\delta\phi}{G(\phi)}[1+\frac{\Gamma}{4H}+([\gamma-1]+\Pi\frac{\zeta_{,\rho}}{\zeta})\frac{\dot{\phi}\Gamma'}{12\gamma H^2}][1+\frac{V}{\lambda}],
\end{eqnarray}
where
\begin{eqnarray}\label{}
\nonumber
G(\phi)=1-\frac{1}{8H^2}[2\gamma\rho+3\Pi+\frac{\gamma\rho+\Pi}{\gamma}(\Pi\frac{\zeta_{,\rho}}{\zeta}-1)](1+\frac{V}{\lambda}).
\end{eqnarray}
In the above equations, for $\Pi\rightarrow 0$ and $\gamma=\frac{4}{3}$ case, we obtain the perturbation variable $\Phi$ of warm inflation model without viscous pressure effect on the brane \cite{6-f} (In this case, we find, $G(\phi)\rightarrow 1,$ because of the inequality,  $\frac{\rho_{\gamma}}{V}\ll 1.$).
By taking scalar field $\phi$ as an independent variable in place of cosmic time $t$, the equation (\ref{22}) could be solved. Using Eq.(\ref{8}), we find
\begin{eqnarray}\label{26}
(3H+\Gamma)\frac{d}{dt}=(3H+\Gamma)\dot{\phi}\frac{d}{d\phi}=-V'\frac{d}{d\phi}.
\end{eqnarray}
From above equation, Eq.(\ref{22}) and Eq.(\ref{25}), the expression $\frac{(\delta\phi)'}{\delta\phi},$ is obtained
\begin{eqnarray}\label{27}
\frac{(\delta\phi)'}{\delta\phi}=\frac{V''}{V'}+\dot{\phi}\frac{\Gamma'}{V'}+\frac{\dot{\phi}}{2HGV'}[\dot{\phi}\Gamma+2V']~~~~~~~~~~~~\\
\nonumber
\times[1+\frac{\Gamma}{4H}+([\gamma-1]+\Pi\frac{\zeta_{,\rho_{\gamma}}}{\zeta})\frac{\dot{\phi}\Gamma'}{12\gamma H^2}][1+\frac{V}{\lambda}].
\end{eqnarray}
We will return to the above relation soon. Following Refs.\cite{1-m}, \cite{6-f} and \cite{10-f},  we introduce auxiliary function $\chi$ as
\begin{equation}\label{28}
\chi=\frac{\delta\phi}{ V'}\exp[\int\frac{1}{3H+\Gamma}\Gamma'd\phi].
\end{equation}
From above definition, we have
\begin{eqnarray}\label{29}
\frac{\chi'}{\chi}=\frac{(\delta\phi)'}{\delta\phi}-\frac{ V''}{ V'}+\frac{\Gamma'}{3H+\Gamma}.
\end{eqnarray}
Using above equation and Eq.(\ref{27}), we find
\begin{eqnarray}\label{30}
\frac{\chi'}{\chi}=\frac{\dot{\phi}}{2HGV'}[\dot{\phi}\Gamma+2V']
[1+\frac{\Gamma}{4H}+([\gamma-1]+\Pi\frac{\zeta_{,\rho_{\gamma}}}{\zeta})\frac{\dot{\phi}\Gamma'}{12\gamma H^2}][1+\frac{V}{\lambda}].
\end{eqnarray}
We could rewrite this equation, using Eqs.(\ref{7}) and (\ref{8})
\begin{eqnarray}\label{31}
\frac{\chi'}{\chi}=
-\frac{3}{8G}\frac{(\Gamma+6H)}{(\Gamma+3H)^2}[\Gamma+4H+([\gamma-1]+\frac{\Pi\zeta_{,\rho_{\gamma}}}{\zeta})\\
\nonumber
\times\frac{V'\Gamma'}{3\gamma H(3H+\Gamma)}]\frac{V'}{V}\frac{(1+\frac{V}{\lambda})}{1+\frac{V}{2\lambda}}.~~~~~~~~~~~~~~~
\end{eqnarray}
A solution for the above equation is
\begin{eqnarray}\label{32}
\chi(\phi)=C\exp(-\int d\phi\frac{3}{8G}\frac{(\Gamma+6H)}{(\Gamma+3H)^2}[\Gamma+4H-([\gamma-1]+\frac{\Pi\zeta_{,\rho_{\gamma}}}{\zeta})\\
\nonumber
\times\frac{V'\Gamma'}{3\gamma H(3H+\Gamma)}]\frac{V'}{V}\frac{(1+\frac{V}{\lambda})}{1+\frac{V}{2\lambda}}),~~~~~~~~~~~~~~~
\end{eqnarray}
where $C$ is integration constant. From above equation and Eq.(\ref{29}), we may find small change of variable $\delta\phi$
\begin{equation}\label{33}
\delta\phi=C V'\exp(\Im(\phi)),
\end{equation}
where
\begin{eqnarray}\label{34}
\Im(\phi)=-\int d\phi[\frac{\Gamma'}{3H+\Gamma}+(\frac{3}{8G}\frac{(\Gamma+6H)}{(\Gamma+3H)^2}[\Gamma+4H-([\gamma-1]+\frac{\Pi\zeta_{,\rho_{\gamma}}}{\zeta})\\
\nonumber
\times\frac{V'\Gamma'}{3\gamma H(3H+\Gamma)}]\frac{V'}{V}\frac{(1+\frac{V}{\lambda})}{1+\frac{V}{2\lambda}})].~~~~~~~~~~~~~~~~~~~~~~~~~~~~~~~~~~~~~~~~~
\end{eqnarray}
Perturbed matter fields of our model are inflaton $\delta\phi$, radiation $\delta\rho$ and velocity $k^{-1}(P+\rho)v_{,i}$. We can explain the cosmological perturbations in terms of gauge-invariant variables. These variables are important for development of the perturbation after the end of the inflation period.  Curvature perturbation $\mathfrak{R},$ and entropy perturbation $e,$ are important gauge-invariant variables which are defied by the perturbed matter fields as a following \cite{nn-2,nn-3}
\begin{eqnarray}\label{}
\mathfrak{R}=\Phi-k^{-1}aHv,~~~~~~~\\
\nonumber
e=\delta P-c_s^2\delta\rho_{\gamma},~~~~~~
\end{eqnarray}
where $c_s^2=\frac{\dot{P}}{\dot{\rho_{\gamma}}}$, and
\begin{eqnarray}\label{}
\Phi\simeq\frac{C\dot{\phi}}{2H}\frac{V'}{G(\phi)}[1+\frac{\Gamma}{4H}+([\gamma-1]+\Pi\frac{\zeta_{,\rho_{\gamma}}}{\zeta})\frac{\dot{\phi}\Gamma'}{12\gamma H^2}][1+\frac{V}{\lambda}]\exp(\Im(\phi).
\end{eqnarray}
In large scale limit, where $k\ll aH$, and in slow-roll limit, the curvature perturbation is given by
\begin{equation}\label{}
 \mathfrak{R}\sim C,
\end{equation}
and the entropy perturbation vanishes \cite{nn-3}.
We can find the density perturbation amplitude by using the above equation and Eq.(\ref{33})  \cite{12-f}
\begin{eqnarray}\label{35}
P_R^{\frac{1}{2}}\sim \mathfrak{R}\sim C,~~~~~~~~~~~~~~~~~~~~~~~~~\\
\nonumber
\delta_H=\frac{2}{5}P_R^{\frac{1}{2}}=\frac{16\pi}{5}\frac{\exp(-\Im(\phi))}{ V'}\delta\phi.
\end{eqnarray}
For high or low energy limit ($V\gg\lambda$ or $V\ll \lambda$), by inserting $\Gamma=0$ and $\Pi=0$, the above equation reduces to $\delta_{H}\simeq\frac{H}{\dot{\phi}}\delta\phi$ which agrees with the density perturbation in cool inflation model \cite{1-i}. In the warm inflation model the fluctuations of the scalar field in high dissipative regime ($r\gg 1$) may be generated by thermal fluctuation instead of quantum fluctuations \cite{5}. Thermal fluctuation of inflaton is given by
\begin{equation}\label{36}
(\delta\phi)^2\simeq\frac{k_F T_r}{2\pi^2}.
\end{equation}
$T_r$ is temperature of the thermal bath. In high dissipative regime, freeze-out wave number $k_F$ is given by
\begin{equation}\label{36}
\nonumber
k_F=\sqrt{\Gamma H}=H\sqrt{3r}\geq H.
\end{equation}
This wave number corresponds to a scale, at which, the dissipation damps out to thermally excited fluctuations, where inequality, $V''<\Gamma H,$ holds \cite{5}. With the help of the above equation, Eqs.(\ref{35}) and (\ref{12}) in high energy ($V\gg\lambda$) and high dissipative regime ($r\gg 1$), we find
\begin{equation}\label{37}
\delta_H^2=(\frac{16\pi}{15})^2\frac{\exp(-2\tilde{\Im}(\phi))}{H^2r^2\dot{\phi}^2}(\delta\phi)^2,
\end{equation}
or equivalently
\begin{eqnarray}\label{}
\nonumber
\delta_H^2\approx \frac{128}{25}\exp(-2\tilde{\Im}(\phi))\frac{T_r}{\tilde{\epsilon}^{\frac{1}{2}}V^{\frac{1}{2}}V'},
\end{eqnarray}
where
\begin{equation}\label{38}
\tilde{\Im}(\phi)=-\int\{\frac{1}{3Hr}\Gamma'+\frac{3}{4\tilde{G}}[1-([\gamma-1]+\Pi\frac{\zeta_{,\rho}}{\zeta})\frac{\Gamma' V'}{3\gamma H(3Hr)^2}]\frac{V'}{V}\}d\phi,
\end{equation}

\begin{equation}\label{}
\nonumber
\tilde{G}(\phi)=1-\frac{V}{8\lambda H^2}[2\gamma\rho_{\gamma}+3\Pi+\frac{\gamma\rho_{\gamma}+\Pi}{\gamma}(\Pi\frac{\zeta_{,\rho_{\gamma}}}{\zeta}-1)],
\end{equation}
and
\begin{equation}\label{39}
\tilde{\epsilon}=\frac{2\lambda}{r}\frac{1}{V}(\frac{V'}{V})^2.
\end{equation}
An important perturbation parameter is scalar index $n_s$ which
in high dissipative regime is given by
\begin{equation}\label{40}
n_s=1+\frac{d\ln \delta_H^2}{d\ln k}\approx
1+\frac{3}{2}\tilde{\eta}-\frac{1}{2}\tilde{\epsilon} -\tilde{\epsilon}(\frac{rV\tilde{\epsilon}}{2\lambda})[\frac{r'}{4r}-2\tilde{\Im}'(\phi)],
\end{equation}
where
\begin{equation}\label{41}
\tilde{\eta}=-\frac{\lambda^2}{V^2}\tilde{\epsilon}+\frac{2\lambda}{rV}(\frac{V''}{V}).
\end{equation}
In Eq.(\ref{40}), we have used a relation between small change of
the number of e-folds and interval in wave number ($dN=-d\ln k$).
Running of the scalar spectral index may be found as
\begin{eqnarray}\label{42}
\alpha_s=\frac{dn_s}{d\ln k}=-\frac{dn_s}{dN}=-\frac{d\phi}{dN}\frac{dn_s}{d\phi}=\frac{2V'\lambda}{rV^2}n_s'.
\end{eqnarray}
This parameter is one of the interesting cosmological
perturbation parameters which is approximately $-0.038$, by using
WMAP observational data \cite{6}. During inflation epoch,
there are two independent components of gravitational waves
($h_{\times +}$) with action of massless scalar field which are
produced by the generation of tensor perturbations. The amplitude
of tensor perturbation is given by
\begin{eqnarray}\label{43}
A_g^2=2(\frac{H}{2\pi})^2\coth[\frac{k}{2T}]=\frac{1}{2\pi^2}\frac{V^2}{6\lambda}\coth[\frac{k}{2T}]=\frac{V^2}{12\pi^2\lambda}\coth[\frac{k}{2T}],
\end{eqnarray}
where, temperature $T$ in extra factor $\coth[\frac{k}{2T}]$
denotes the temperature of  thermal background of
gravitational waves \cite{7}. Spectral index $n_g$ may be found as
\begin{eqnarray}\label{44}
n_g=\frac{d}{d\ln k}(\ln [\frac{A_g^2}{\coth(\frac{k}{2T})}])\simeq-2\tilde{\epsilon},
\end{eqnarray}
where, $A_g\propto k^{n_g}\coth[\frac{k}{2T}]$ \cite{7}.  Using Eqs. (\ref{37})  and
(\ref{43}) we write the tensor-scalar ratio in high dissipative
regime
\begin{eqnarray}\label{45}
R(k)=\frac{A_g^2}{P_R}|_{k=k_{0}}=\frac{2}{3\pi^2\lambda}\frac{\tilde{\epsilon}^{\frac{1}{2}}V^{\frac{5}{2}}V'}{T_r}\exp(2\tilde{\Im}(\phi))\coth[\frac{k}{2T}],
\end{eqnarray}
where $k_{0}$ is referred  to pivot point \cite{7}. An upper bound for this parameter is obtained
by using Planck data, $R<0.11$ \cite{planck}.

We note that, the $\Im(\phi)$ factor (\ref{34})  which is found  in perturbation parameters (\ref{37}), (\ref{40}), (\ref{42}) and (\ref{45}) in high energy limit ($V\gg\lambda$), for warm-viscous inflation model on the brane has an important difference with the same factor which was obtained for usual warm-viscous   inflation model \cite{2-ne}
\begin{eqnarray}\label{}
\nonumber
\Im(\phi)=-\int d\phi[\frac{\Gamma'}{3H+\Gamma}+(\frac{3}{8G}\frac{(\Gamma+6H)}{(\Gamma+3H)^2}[\Gamma+4H-([\gamma-1]+\frac{\Pi\zeta_{,\rho_{\gamma}}}{\zeta})\\
\nonumber
\times\frac{V'\Gamma'}{3\gamma H(3H+\Gamma)}]\frac{V'}{V})].~~~~~~~~~~~~~~~~~~~~~~~~~~~~~~~~~~~~~~~~~
\end{eqnarray}
The density square term in the effective  Einstein equation on the brane leads to this difference.
Therefore, the perturbation parameters which may be found by WMAP observational data, for our model on the brane, are modified due to the effect of this term. In the other hand, the slow-roll parameters (\ref{9}) and (\ref{10}) which are derived in background level, are modified because of the density square term in modified Friedmann equation (\ref{6}). The slow-roll parameters are  appeared in the perturbation parameters  (\ref{40}), (\ref{42}), (\ref{44}) and (\ref{45}). So, from above discussion, we know the density square term in the effective Einstein equation on the brane gives the significant contributions to the observable parameters, $P_R$, $R$, $n_s$ and $\alpha_s$. Also, the different observable perturbation parameters for models of non-viscous  warm inflation and non-viscous warm inflation  on the brane may be found in Ref.\cite{9-f} and \cite{6-f}.

\section{Chaotic  potential }
In the present section we will study the  warm-viscous  inflation model on the brane  using chaotic potential
\begin{equation}\label{46}
V(\phi)=\frac{1}{2}m^2\phi^2,
\end{equation}
where $m>0$ is a constant free parameter. We restrict our study in high dissipative and high energy limits for two cases: 1- $\Gamma$ and $\zeta$ are constant parameters. 2- $\Gamma$ is a function of scalar field $\phi$ and $\zeta$ is a function of energy density  $\rho$ of imperfect fluid.
\subsection{$\Gamma=\Gamma_0$, $\zeta=\zeta_0$ case}
We take dissipative coefficient $\Gamma=\Gamma_0$ and bulk viscosity coefficient $\zeta=\zeta_0$ as constant parameters. The Eq.(\ref{7}) in high dissipative regime reduces to
\begin{equation}\label{47}
\dot{\phi}=-\frac{V'}{3Hr}.
\end{equation}
The inflaton field in term of cosmic time has the form
\begin{equation}\label{48}
\phi=\phi_0\exp(-\frac{m^2}{\Gamma_0}t).
\end{equation}
Using Eq.(\ref{6}) and chaotic potential (\ref{46}), the Hubble parameter becomes
\begin{equation}\label{49}
H(\phi)=\frac{m^2}{2\sqrt{6\lambda}}\phi^2,
\end{equation}
and the parameter $r=\frac{\Gamma}{3H}$ is given by
\begin{equation}\label{50}
r=\frac{2\sqrt{6\lambda}\Gamma_0}{3m^2\phi^2}.
\end{equation}
The slow-roll parameter $\tilde{\epsilon}$ in the present case is obtained, using Eq.(\ref{39})
\begin{equation}\label{51}
\tilde{\epsilon}=\frac{2\lambda}{r}\frac{1}{V}(\frac{V'}{V})^2=\frac{4\sqrt{6\lambda}}{\Gamma_0\phi^2}.
\end{equation}
With the help of  Eqs.(\ref{37}) and (\ref{45}), the scalar power spectrum and tensor-scalar ratio result to be
\begin{equation}\label{52}
P_R=16(\frac{V(\phi_0)}{m^2})^{\frac{3}{2+\frac{3}{4}\zeta_0\sqrt{\frac{6}{\lambda}}}}
\frac{\Gamma_0^{\frac{1}{2}}T_r}{m^2(6\lambda)^{\frac{1}{4}}V(\phi_0)^{\frac{3}{2}}},
\end{equation}
and
\begin{equation}\label{53}
R=\frac{4\sqrt[4]{6}}{3\pi^2\lambda^{\frac{3}{4}}}
(\frac{V(\phi_0)}{m^2})^{\frac{-3}{2+\frac{3}{4}\zeta_0\sqrt{\frac{6}{\lambda}}}}
\frac{m^2V(\phi_0)^{\frac{7}{2}}}{\Gamma_0^{\frac{1}{2}}T_r}\coth[\frac{k}{2T}],
\end{equation}
respectively, where the subscript zero $0$ denotes the time, when  the perturbation was leaving the horizon. These parameters may be found from WMAP and Planck observational data \cite{6,planck}. Using these data ($R<0.11$, $P_R=2.28\times 10^{-9}$), we find an upper bound for $V(\phi_0)$
\begin{equation}\label{56}
\nonumber
V(\phi_0)<0.28\times 10^{-12}
\end{equation}
The energy density of imperfect fluid in term of energy density of inflaton from Eq.(\ref{8}) is given by
\begin{equation}\label{54}
\rho_{\gamma}=\frac{1}{\gamma}(\frac{2\sqrt{6\lambda}m^2}{3\Gamma_0}+\frac{2\zeta_0 m^2}{2\sqrt{6\lambda}}\phi^2)=\frac{\sqrt{6\lambda}}{\gamma}(\frac{2m^2}{3\Gamma_0}+\frac{\zeta_0}{2\lambda}\rho_{\phi}).
\end{equation}
For this example, the  entropy density in terms of cosmic time may be derived from Eqs. (\ref{48}),(\ref{54})
\begin{equation}\label{}
Ts=\frac{\sqrt{6\lambda}m^2}{2\Gamma_0}+\frac{3\zeta_0 m^2\phi_0}{4\sqrt{6\lambda}}\exp(-\frac{m^2}{\Gamma_0}t).
\end{equation}
In FIG.1,  we plot the entropy density in terms of cosmic time. It can be seen that, the entropy density increases by the bulk viscous effect \cite{mm-1}.
The number of e-folds at the end of inflation is found, using Eq.(\ref{14})
\begin{equation}\label{55}
N(\phi)=-\frac{\Gamma_0}{2\sqrt{6\lambda}}(\phi_f^2-\phi_i^2),
\end{equation}
or equivalently
\begin{equation}\label{56}
N(\phi)=\frac{\Gamma_0}{m^2\sqrt{6\lambda}}(V_i-V_f),
\end{equation}
where $V_i>V_f$.
At the end of inflation ($\tilde{\epsilon}\simeq 1$), the inflaton $\phi$ results to be
\begin{equation}\label{57}
\phi_f^2=\frac{4\sqrt{6\lambda}}{\Gamma_0}.
\end{equation}
From above equation and Eq.(\ref{55}), initial inflaton $\phi_i$ may be found in term of the number of e-folds
\begin{equation}\label{58}
\phi_i^2=(\frac{N_{total}}{2}+1)\phi_f^2=\frac{4\sqrt{6\lambda}}{\Gamma_0}(\frac{N_{total}}{2}+1).
\end{equation}
Inserting $N_{total}\approx 60,$ we get $\phi_i\approx 17.43\frac{\lambda^{\frac{1}{4}}}{\Gamma_0}$.
From Eq.(\ref{50}), $r$ parameter at the beginning of inflationary era has the form
\begin{equation}\label{59}
r(\phi=\phi_i)=r_i=\frac{1}{186}(\frac{\Gamma_0}{m})^2.
\end{equation}
Therefore, in high dissipative regime $r_i\gg 1,$   we have $\Gamma_0\gg \sqrt{186} m$.
\subsection{$\Gamma=\Gamma(\phi)$, $\zeta=\zeta(\rho_{\gamma})$ case}
Now we assume $\zeta=\zeta(\rho_{\gamma})=\zeta_1\rho,$ and $\Gamma=\Gamma(\phi)=\alpha V(\phi)=\alpha\frac{m^2\phi^2}{2}$, where $\alpha$ and $\zeta_1$ are positive constants. By using chaotic potential (\ref{46}), Hubble parameter $H$, $r$ parameter and slow-roll parameter $\tilde{\epsilon}$  have the forms
\begin{eqnarray}\label{60}
H(\phi)=\frac{m^2\phi^2}{2\sqrt{6\lambda}},~~~~~~~~~~~~~~~~~~r=\sqrt{\frac{2}{3}\lambda}\alpha,~~~~~~~~~~~~~~~~~
\tilde{\epsilon}=\frac{8\sqrt{6\lambda}}{\alpha m^2\phi^4},
\end{eqnarray}
respectively. In high dissipative regime $r\gg 1,$ we have $\alpha\gg\sqrt{\frac{3}{2\lambda}}$.  Using Eq.(\ref{47}), we find the scalar field $\phi$ in term of cosmic time
\begin{equation}\label{61}
\phi^2(t)=-\frac{4}{\alpha}t+\phi_i^2.
\end{equation}
Using Eq.(\ref{11}), the energy density of imperfect fluid $\rho$ in term of the inflaton energy density $\rho_{\phi}$ is given by the expression
\begin{equation}\label{62}
\rho_{\gamma}=\frac{16\sqrt{6\lambda}}{3\gamma\alpha}\frac{1}{(1-\frac{3\zeta_1 m^2\phi^2}{2\sqrt{6\lambda}\gamma})}=\frac{16\sqrt{6\lambda}}{3\gamma\alpha}\frac{1}{(1-\frac{3\zeta_1}{\sqrt{6\lambda}\gamma}\rho_{\phi})}.
\end{equation}
We can find the entropy density $s$ in terms of cosmic time
\begin{equation}\label{}
Ts=\frac{16\sqrt{6\lambda}}{3\gamma\alpha}\frac{1}{(1-\frac{6\zeta_1 m^2}{\sqrt{6\lambda}\gamma\alpha}t-\frac{3\zeta_1 m^2}{2\sqrt{6\lambda}\gamma}\phi_i)}.
\end{equation}
The entropy density and energy density of our model in this case increase by the bulk viscosity effect (see FIG.2).
Number of e-folds in this case is obtained by using Eq.(\ref{14})
\begin{equation}\label{63}
N(\phi)=-\frac{\alpha m^2}{4\sqrt{6\lambda}}(\phi_f^4-\phi_i^4),
\end{equation}
or equivalently
\begin{equation}\label{64}
N(\phi)=-\frac{\alpha}{m^2\sqrt{6\lambda}}(V_f^2-V_i^2),
\end{equation}
where $V_i>V_f$. From Eq.(\ref{61}) the scalar field at the end of inflation where $\tilde{\epsilon}\simeq 1$, becomes
\begin{equation}\label{65}
\phi_f^4=\frac{8\sqrt{6\lambda}}{\alpha m^2}.
\end{equation}
Therefore inflaton $\phi_i$ in term of number of e-folds is given by
\begin{equation}\label{66}
\phi_i=(\frac{N_{total}}{2}+1)^{\frac{1}{4}}(\frac{8\sqrt{6\lambda}}{\alpha m^2})^{\frac{1}{4}}=(\frac{N_{total}}{2}+1)^{\frac{1}{4}}\phi_f.
\end{equation}
By using  Eqs.(\ref{37}) and (\ref{45}) the scalar power spectrum and tensor-scalar ratio result to be
\begin{equation}\label{67}
P=\frac{16 T_r\sqrt{\alpha}}{m^2(6\lambda)^{\frac{1}{4}}}V(\phi_0)^{\frac{7}{2}}\exp(\frac{A}{V(\phi_0)^2}+\frac{B}{V(\phi_0)}),
\end{equation}
and
\begin{equation}\label{68}
R=\frac{4}{3\pi^2\lambda}\frac{m^2(6\lambda)^{\frac{1}{4}}}{T_r\sqrt{\alpha}}V(\phi_0)^{-\frac{3}{2}}\exp(-\frac{A}{V(\phi_0)^2}-\frac{B}{V(\phi_0)})\coth[\frac{k}{2T}],
\end{equation}
respectively, where $A=-3(\gamma-1)\frac{\sqrt{6\lambda}m^2}{3\alpha\gamma}$, and $B=\frac{\zeta_1^2 m^2}{\alpha\gamma}$. We may restrict  these parameters, using WMAP9 and Planck observational data \cite{6}.
Based on these data, an upper bound for $V(\phi_0)$ may be found
\begin{equation}\label{}
\nonumber
V(\phi_0)<1.9\times 10^{-2}.
\end{equation}
In the above equation we have used these data: $R<0.11$, $P_R=2.28\times 10^{-9}$ \cite{6}.
Using WMAP9 data, $P_R(k_0)=\frac{25}{4}\delta_H^2\simeq 2.28\times 10^{-9}$, BICEP2 data $R(k_0)\simeq 0.11,$ \cite{BICEP2} and the characteristic of warm inflation, $T>H$ \cite{3}, we may restrict the values of temperature
($T_r>7.73\times 10^{-5}M_4$) using Eqs.(\ref{37}), (\ref{45}), or the corresponding equations (\ref{52}), (\ref{53}), (\ref{67}), (\ref{68}),
in our coupled examples, (see FIG.3). We have chosen $k_0=0.002 Mpc^{-1}$ and $T\simeq T_r$. Note that, because of the bulk viscous pressure,
the radiation energy density in our model increases. Therefore the minimum value of temperature for our model ($7.73\times 10^{-5}M_4$)
is bigger than the minimum value of temperature ($3.42\times 10^{-6}M_4$ ) for the model without the viscous pressure effects \cite{6-f}. By using BICEP2 data, we have presented a new minimum of $T$ (See for example \cite{end}).
\begin{figure}[h]
\centering
  \includegraphics[width=10cm]{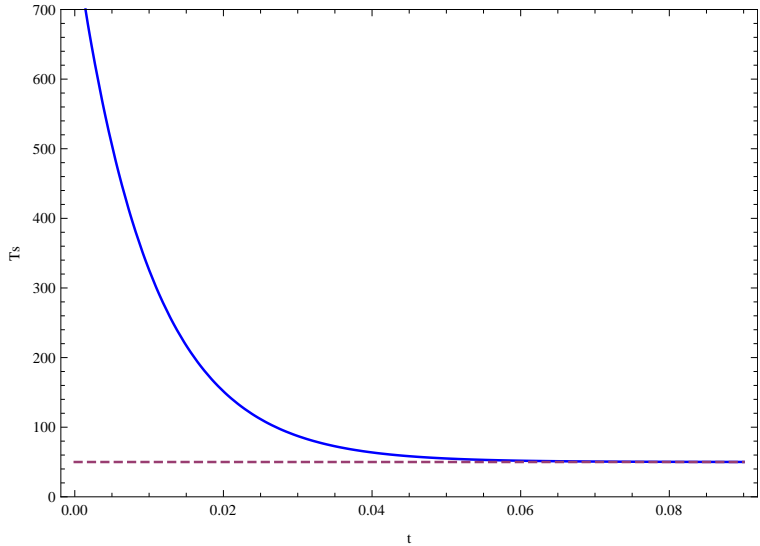}
  \caption{We plot the entropy density $s$ in terms of cosmic time $t$ where, $\Pi=0$ by dashed curve and $\Pi=-3\zeta_0 H$ by blue curve ($T=5.47\times 10^{-5} M_4$, $\Gamma=\Gamma_0=8\pi\times 10^{-3}, \zeta_0=(8\pi)^{-1}\times10^{3}, m^2=8\pi\times 10^{-1},\phi_0^2=10, \gamma=\frac{4}{3}, \sqrt{6\lambda}=1$)}
 \label{fig:F1}
\end{figure}

\begin{figure}[h]
\centering
  \includegraphics[width=10cm]{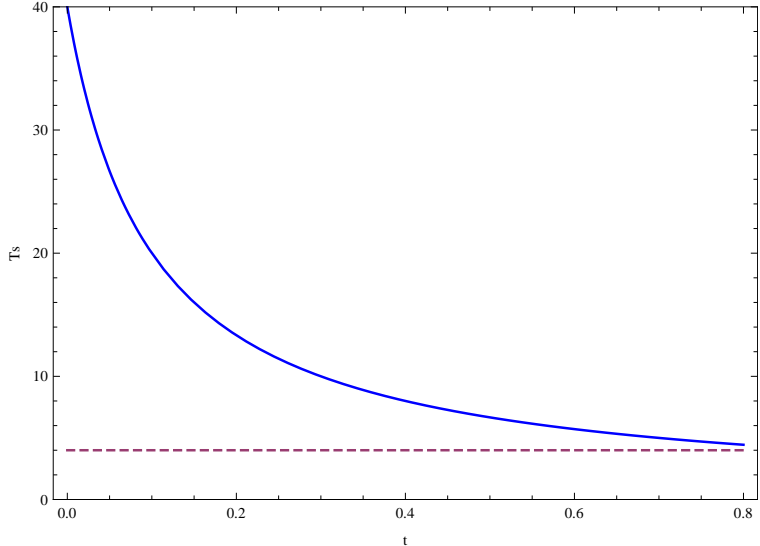}
  \caption{We plot the entropy density $s$ in terms of cosmic time $t$ where, $\Pi=0$ by dashed line and $\Pi=-3\zeta_1\rho H$ by blue curve ($T=5.47\times 10^{-5}M_4, \phi_0=3.6, \sqrt{6\lambda}=1, \gamma=\frac{4}{3}, \alpha=1,\zeta_1=3.9\times10^{-10},m=2.38\times 10^4$)}
 \label{fig:F2}
\end{figure}

\begin{figure}[h]
\centering
  \includegraphics[width=10cm]{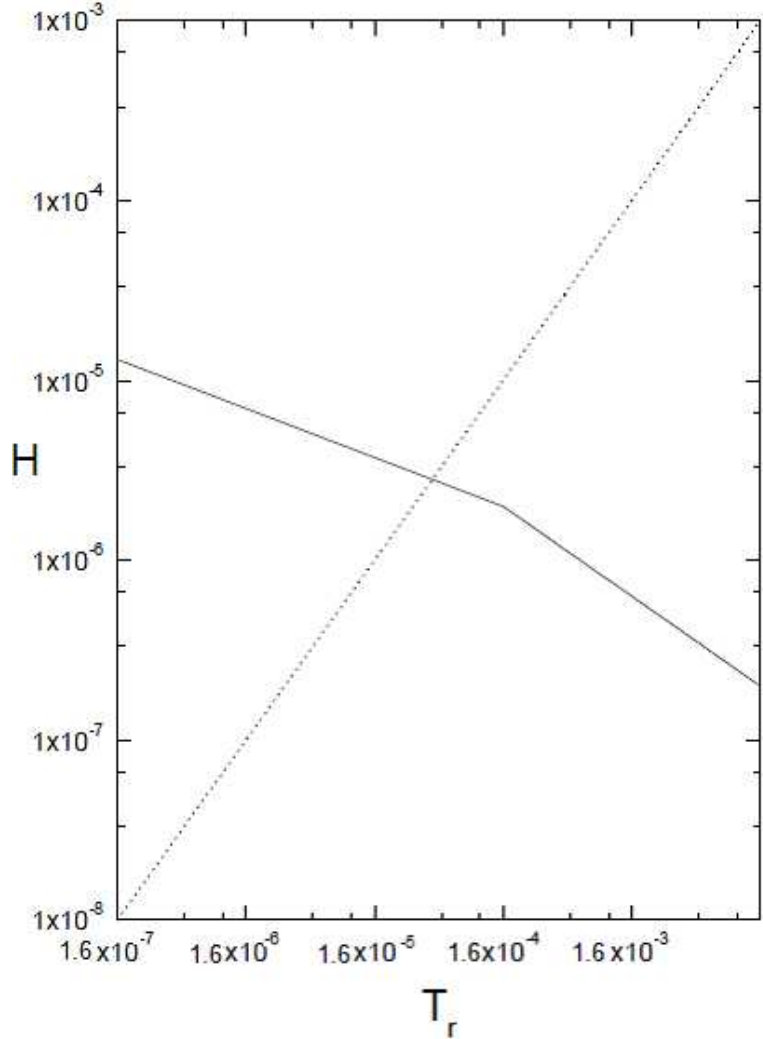}
  \caption{In this graph we plot the Hubble parameter $H$ in term of the temperature $T_r$. We can find the minimum amount of temperature $T_r=5.47\times 10^{-5}$ in order to have the necessary condition for warm inflation model ($T_r>H$). }
 \label{fig:F3}
\end{figure}
\section{Conclusion}
 One  problem of the
(cold)inflation theory, is how to attach the universe to the end of the
inflation period. One of the solutions of this problem is the
study of inflation in the context of warm inflation \cite{3}. In
this model, radiation is produced during inflation period where its
energy density is kept nearly constant. This is phenomenologically
fulfilled by introducing the dissipation term $\Gamma$. The study
of warm inflation model with viscous pressure is an extension of warm inflation model
where instead of radiation field we have radiation-matter fluid. In this article we have considered warm
inflationary universe  model with viscous pressure on the brane. In the slow-roll approximation, the general relation between energy density of radiation-matter mixture  and energy density of scalar field is found. In longitudinal gauge and slow-roll limit the explicit expressions
for the tensor-scalar ratio $R,$ scalar spectrum $P_R,$ scalar index
$n_s,$ and its running $\alpha_s$ have been obtained. We have
developed our specific model by chaotic potential in two cases: 1- $\Gamma$ and $\zeta$ are constant parameters. 2- $\Gamma$ is a function of scalar field $\phi$ and $\zeta$ is a function of energy density of imperfect fluid. In these two
cases  we have found perturbation parameters and constrained these
parameters by WMAP9, BICEP2 and Planck observational data.

\end{document}